\documentstyle[prb,aps,multicol,epsfig]{revtex}
\renewcommand{\narrowtext}{\begin{multicols}{2}
\global\columnwidth20.5pc}
\renewcommand{\widetext}{\end{multicols}
\global\columnwidth42.5pc} \multicolsep = 8pt plus 4pt minus 3pt

\newcommand{\twobeone}{\widetext\vskip.6pc \noindent \vrule
width3.375in height.2pt depth.2pt \vrule depth0em height1em\hfill
\vskip.6pc }
\newcommand{\onebetwo}{
\vskip.6pc \indent
 \hfill\vrule depth1em height0pt \vrule width3.375in height.2pt depth.2pt
\vskip.6pc \narrowtext \noindent}

\draft

\title{Hysteresis effect in $\nu=1$ quantum Hall system under
periodic electrostatic modulation}

\author {Ming-Che Chang${}^*$ and Min-Fong Yang${}^\dagger$
}

\address{
${}^*$ Department of Physics, National Taiwan Normal University,
Taipei, Taiwan
\\${}^\dagger$ Department of Physics, Tunghai University,
Taichung, Taiwan }

\date{\today}

\begin{document}

\maketitle

\begin{abstract}

The effect of a one-dimensional periodic electrostatic modulation
on quantum Hall systems with filling factor $\nu=1$ is studied. We
propose that, either when the amplitude of the modulation
potential or the tilt angle of the magnetic field is varied, the
system can undergo a first-order phase transition from a fully
spin-polarized homogeneous state to a partially spin-polarized
charge-density-wave state, and show hysteresis behavior of the
spin polarization. This is confirmed by our self-consistent
numerical calculations within the Hartree-Fock approximation.
Finally we suggest that the $\nu=1/3$ fractional quantum Hall
state may also show similar hysteresis behavior in the presence of
a periodic potential modulation.

\end{abstract}


\narrowtext


The discovery of the integer and fractional quantum Hall (QH)
effects offered invaluable tools to study quantum phase
transitions in low dimensions.\cite{reviews} For example, the
translational-invariant QH phases can exhibit novel forms of
two-dimensional ferromagnetism, and show interesting magnetic
transitions. Recently, the hysteresis phenomena reminiscent of
conventional ferromagnetic materials are discovered in many QH
systems. For instance, in close proximity to the critical pressure
necessary for the transition from the spin-polarized state to the
spin-unpolarized state, a hysteretic evolution of the
magnetoresistance is observed in the vicinity of {\it
even-numerator} fractional quantum Hall states.\cite{Cho_et_al}
Furthermore, by applying a gate bias, hysteresis behavior of the
longitudinal resistivity is observed in a {\it wide} quantum well
at {\it even-integer} QH states.\cite{Piazza_et_al} The physical
origin of these observed hysteresis may be associated with the
crossing of Landau levels for electrons (or for composite fermions
in the case of the fractional QH states) with different spin
polarizations.\cite{Jungwirth_et_al}

Besides the hydrostatic pressure and the gate bias, modern
techniques allow us to introduce other external perturbations to
QH systems, such as a lateral periodic electrostatic potential
and/or a periodic magnetic field. QH systems under periodic
modulations, either one-dimensional\cite{modulation1} or
two-dimensional,\cite{modulation2} have been studied to a great
extent. Recently, Manolescu and Gudmundsson have theoretically
studied a QH system at {\it non-integer} filling factor $\nu$
(say, $\nu\simeq 2.42$ in Ref.~\onlinecite{Manolescu00:7858})
under periodic
modulations.\cite{Manolescu99:5426,Manolescu00:7858} By varying
either the amplitude of the modulation or the tilt angle of the
magnetic field with its normal component $B_0$ being fixed, they
found a hysteretic evolution of the ground state due to the
combined effects of the external potential and the exchange
interaction. It is interesting to study if similar hysteresis
effects can be found in other situations.

In this paper, we consider the $\nu=1$ QH system in a single layer
under a one-dimensional periodic modulation. For the system
without external periodic modulation, it is well-known that, even
when the Zeeman splitting is negligible, because of the strong
exchange interaction among electrons, the ground state is a
homogeneous, fully spin-polarized incompressible
liquid.\cite{Ando} However, Bychkov {\it et al.} found that, when
the mixing between the Landau levels is neglected, the presence of
a one-dimensional periodic modulation could diminish the
excitation energy of spin excitons with {\it finite} momentum to
zero.\cite{Bychkov} Due to these gapless spin excitons, one may
expect that the system can transit from a fully spin-polarized
homogeneous state to a partially spin-polarized
charge-density-wave (CDW) state. While Bychkov {\it et al.}
suggested this possibility, they did not pursue this issue
further, because their approach is applicable only prior to the
occurrence of this instability.

We notice that the proposed instability is very likely a
first-order phase transition accompanying hysteresis phenomena.
This can be understood as follows. First, this instability is
related to the crossing of the single-particle Landau bands with
different spin polarizations. Similar picture has been used to
explain the observed hysteresis due to the application of the
hydrostatic pressure or the gate bias.\cite{Jungwirth_et_al} It
indicates that the proposed transition could be first-order and
show a hysteresis effect.
Second, it is expected that, by varying the modulation amplitude,
the evolution of the ground state could be history-dependent. If
the system is originally
in the uniform fully spin-polarized state, 
as the modulation amplitude increases, in order to minimize the
external potential energy, electrons tend to accumulate at the
local minima of the external modulation potential. However, this
process must accompany with spin flipping and therefore will cost
exchange energy among the spin-up electrons. When the cost in
exchange energy is larger than the reduction in the external
potential energy, the uniform fully spin-polarized state is
maintained even under a modest modulation. 
The rigidity of the uniform phase breaks down only when
the modulation strength exceeds a critical value, and a sudden
transition to a partially spin-polarized state ensues. 
In contrast, once the system is in a partially spin-polarized
state (now electrons with both spin orientations are present),
when the modulation strength is reduced (the local minima become
shallower), minority-spin (spin-down) electrons will be squeezed
out of the potential minima one by one. Note that, in this case,
when a minority-spin electron is pushed back to the nearby
unoccupied majority-spin (spin-up) state due to the reduction of
the modulation amplitude, the gain in exchange energy among the
spin-up electrons is roughly the same order as the cost among the
spin-down electrons. That is, the effect of the exchange
interaction is minor in the present case. Therefore, both the
evolution of the ground state and the change of the population are
mainly determined by the external modulation potential, and they
change gradually as the modulation is varied slightly. Hence the
fully spin-polarized state will not immediately be restored even
though the modulation strength is decreased below the previous
critical
value. 
Only when the modulation strength is sufficiently small, such that
the difference of the populations between the spin-up and
spin-down states exceeds a critical amount, will the spin gap be
abruptly
amplified, and the fully polarized state be recovered. 
This explains why a hysteresis effect appears in the present
system. The above argument is confirmed by our calculations within
the Hartree-Fock approximation (HFA). 
We find that, either by changing the amplitude of the modulation
potential or the tilt angle of the magnetic field, hysteresis
behavior of the spin polarization will occur in the modulated
$\nu=1$ QH system.


By using the Landau gauge and neglecting the Landau level mixing,
which is valid when the cyclotron energy $\hbar\omega_c$ is much
larger than the typical Coulomb energy, the Hamiltonian of a
modulated QH system under a strong perpendicular magnetic field $B_0$
can be expressed in the form
\begin{eqnarray}
H&=& \sum_{X, \alpha} ( -\frac{\alpha}{2} \Delta_z^{(0)} - \mu )
c_{X, \alpha}^{\dagger} c_{X, \alpha} + H_M + H_C,
\label{Hamiltonian} \\ H_M &=& \sum_{X, \alpha} V_0 e^{-(G_0 l)^2/4}
\cos\left( G_0 X \right) c_{X, \alpha}^{\dagger} c_{X, \alpha} \; ,
\label{H_M}
\end{eqnarray}
where $c_{X, \alpha}^{\dagger}$ is the creation operator of an
electron in the lowest Landau level (LLL) with a guiding center
coordinate $X$ and spin $\alpha$ ($\alpha =\pm 1$).
$\Delta_z^{(0)}$, $\mu$, and $l \equiv \sqrt{\hbar /eB_0}$ are the
Zeeman energy, the chemical potential, and the magnetic length,
respectively.
Here we take $\hbar\omega_c /2$ as the zero of energy. In the
present investigation, we consider the simplest model of a
one-dimensional electrostatic modulation $V(x)=V_0 \cos (G_0 x)$
with a period $a=2\pi/G_0$. Finally, the Coulomb interaction gives
a contribution $H_C$, \twobeone
\begin{eqnarray}
H_C &=& \frac{1}{2} \sum_{\{X_{i}\}, \alpha, \beta} \langle X_1,
X_2|v|X_3, X_4\rangle c_{X_1, \alpha}^{\dagger} c_{X_2,
\beta}^{\dagger} c_{X_3, \beta} c_{X_4, \alpha} \; , \label{H_C} \\
\langle X_1, X_2|v|X_3, X_4\rangle &=& \frac{1}{A} \sum_{\bf q}
v({\bf q}) e^{-(q l)^2/2} e^{iq_{x}(X_1 + X_4)/2} e^{-iq_{x}(X_2 +
X_3)/2}
 \delta_{X_1, X_4 + q_y l^2} \delta_{X_2, X_3 - q_y l^2} \; ,
\end{eqnarray}
\onebetwo where $A$ is the area of the system and $v({\bf q})=2\pi
e^2/\kappa |{\bf q}|$ is the Fourier transform of the Coulomb
potential with $\kappa$ being the dielectric constant.

In the spirit of the HFA, the model Hamiltonian becomes
\begin{equation}
H_{\rm HF} = \sum_{X, \alpha} ( \varepsilon_{X, \alpha} - \mu ) c_{X,
\alpha}^{\dagger} c_{X, \alpha} \; , \label{HFA}
\end{equation}
where the single-particle energies under the HFA are
\begin{eqnarray}\label{HF_energy}
\varepsilon_{X, \alpha} = &-&\frac{\alpha}{2} \Delta_z^{(0)} + V_0
e^{-(G_0 l)^2/4} \cos ( G_0 X ) \nonumber \\ &+& \sum_{G_j}
W_0^{\alpha} (G_j) e^{-i G_j X}
\end{eqnarray}
with $\{G_j = j G_0, j=0, \pm 1, \pm 2 \cdots\}$ being the reciprocal
lattice vectors of the one-dimensional periodic structure. The
Hartree-Fock effective interaction potential $W_0^{\alpha}(G)$ can be
split into a direct [$H_0(G)$] and an exchange [$X_0(G)$]
component\cite{Cote}
\begin{equation}\label{effectivepot}
W_0^{\alpha}(G) = \frac{e^2}{\kappa l} \sum_{\beta} \left[ H_0(G) -
\delta_{\alpha,\beta} X_0(G) \right] \left\langle \rho_0^{\beta}(-G)
\right\rangle
\end{equation}
with
\begin{eqnarray}
H_0(G) &=& \frac{1}{|G|} e^{-(G l)^2/2} \left(
1-\delta_{G,0}\right), \label{hartree}
\\ X_0(G) &=& \sqrt{\frac{\pi}{2}} e^{-(G l)^2/4}
I_0\left[ \frac{(G l)^2}{4} \right], \label{fock} \\ \left\langle
\rho_0^{\beta}(-G) \right\rangle &=& \frac{1}{N_{\varphi}}
\sum_{X} e^{iGX} \langle c_{X, \beta}^{\dagger} c_{X, \beta}
\rangle, \label{rho}
\end{eqnarray}
where $I_0(x)$ is the modified Bessel function and $N_{\varphi}$ is
the Landau level degeneracy. The factor $\left( 1-\delta_{G,0}
\right)$ in the expression of $H_0(G)$ is due to the neutralizing
positive background. From Eq.~(\ref{HF_energy}), we find that the
modulation lifts the degeneracy of the Landau levels and the
resulting single-particle energies have a periodic structure,
$\varepsilon_{X, \alpha}=\varepsilon_{X+a, \alpha}$. Within the HFA,
the thermal expectation value $\langle c_{X, \alpha}^{\dagger} c_{X,
\alpha} \rangle = f(\varepsilon_{X, \alpha} - \mu)$, where $f(x)$ is
the Fermi-Dirac distribution function. The condition of $\nu=1$ is
given by $\left(1/N_{\varphi}\right)\sum_{X, \alpha} \langle c_{X,
\alpha}^{\dagger} c_{X, \alpha} \rangle =1.$
Eqs.~(\ref{HF_energy})-(\ref{rho}), together with this condition,
form the basis of our self-consistent scheme.

Before starting the numerical calculations, we notice that, for
$\nu=1$, the Hamiltonian within the LLL approximation
[Eqs.~(\ref{Hamiltonian})-(\ref{H_C})] is invariant under the
following particle-hole transformation,
\begin{equation}
c_{X, \alpha}^{\dagger} \rightarrow \tilde{c}_{X+a/2, -\alpha} \;\;
\hbox{and}\;\; c_{X, \alpha} \rightarrow \tilde{c}_{X+a/2,
-\alpha}^{\dagger} \; .
\end{equation}
By using this particle-hole symmetry, one leads to an exact
expression of $\mu$ at all temperatures, $\mu = -\epsilon_c/2$,
where $\epsilon_c \equiv \sqrt{\pi/2}\;(e^{2}/\kappa l)$ is the
exchange energy per electron of the uniform fully polarized state.
For simplicity, in the following, the units of length and energy are
taken to be the magnetic length $l$ and $\epsilon_c$, respectively.


\begin{figure}
\centerline{\epsfxsize=7cm \epsfbox{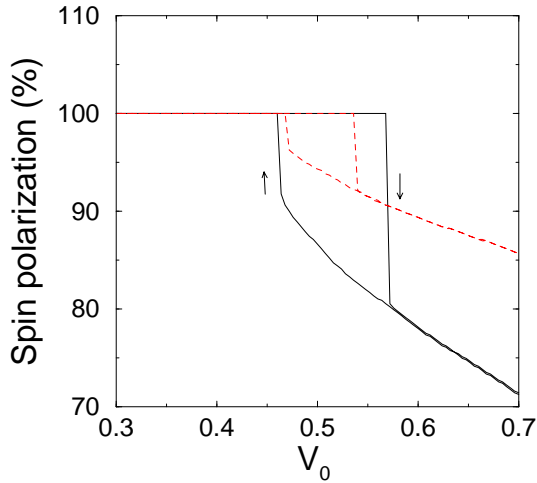} } \caption{Evolution
of the spin polarization for $a=10$ (solid line) and $a=15.7$
(dashed line) at $T=0$.
 }\label{fig1}
\end{figure}

We consider two modulation periods, $a=10$ and $a=15.7$ (i.\ e.\
$G_0=0.4$), with $\Delta_z^{(0)}=0.03$. The period $a=15.7$ is
chosen in order to compare our result with that in
Ref.~\onlinecite{Bychkov}. We begin with the potential amplitude
$V_0=0$, and find the self-consistent solution by iteration,
starting from the fully polarized state. Then the value of $V_0$
is increased slightly and a new self-consistent solution is
obtained by using the previous solution as the initial try. The
modulation amplitude is then changed again and this
self-consistent scheme is repeated. As shown in Fig.~\ref{fig1},
by increasing and then decreasing $V_0$, we obtain hysteresis
behavior of the spin polarization, $\left(1/N_{\varphi}\right)
\sum_{X, \alpha} \alpha \langle c_{X, \alpha}^{\dagger} c_{X,
\alpha} \rangle$. This indicates that the system undergoes a
first-order phase transition from a fully polarized homogeneous
state to a partially polarized CDW state (see the discussions
below),
which corresponds to the sudden drop of the spin polarization in
Fig.~\ref{fig1}. Within the HFA, the critical value of $V_0$ at
which the instability occurs can be estimated as follows. For the
uniform fully polarized state to be the self-consistent solution
at $T=0$, the maximal value of $\varepsilon_{X, +1}$ for this
state must be lower than the chemical potential $\mu =
-\epsilon_c/2$. From Eq.~(\ref{HF_energy}), this gives
\begin{equation}\label{SC_condition}
V_0 \leq \frac{1}{2} \left( \Delta_z^{(0)} + \epsilon_c \right)
e^{(G_0 l)^2/4}.
\end{equation}
Thus the critical value of $V_0$ is given by the right-hand side of
the above inequality.\cite{note2} For $\Delta_z^{(0)}=0.03$, the
critical value is 0.568 (0.538) for $a=10$ ($a=15.7$), which agrees
with the result in Fig.~\ref{fig1}. Moreover, from Fig.~\ref{fig1} we
find that the larger the period $a$ is, the smaller the hysteresis
loop becomes.
Thus our result implies that the hysteresis loop will no longer
exist for a long enough period.

\begin{figure}
\centerline{\epsfxsize=7cm  \epsfbox{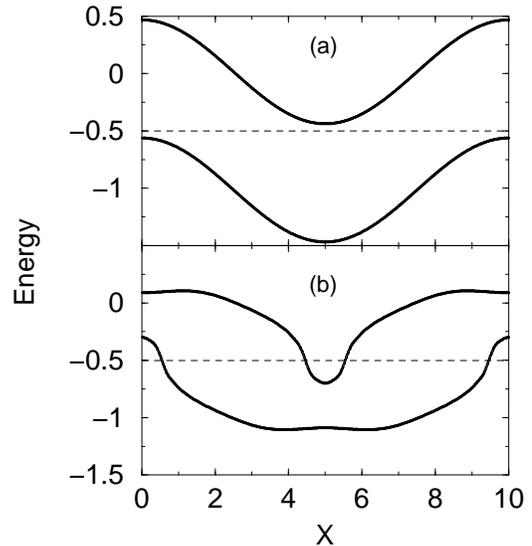} } \caption{The
self-consistent solution of the energy spectra in the first
Brillouin zone, $0\leq X < a$ for $a=10$. The modulation amplitude
$V_0=0.5$ in Fig. 2(a) and $V_0=0.6$ in Fig. 2(b). The horizontal
dashed lines denote the Fermi levels. }\label{fig2}
\end{figure}

To investigate the electron population in the ground state, energy
spectra $\varepsilon_{X, \alpha}$ for two different $V_0$'s with
$a=10$ and $\Delta_z^{(0)}=0.03$ are shown in Fig.~\ref{fig2}. For
$V_0$ smaller than the critical value [Fig.~\ref{fig2}(a)], the
spin splitting is amplified by the exchange energy $\epsilon_c$,
and there is no overlap between the spin-split bands. Therefore,
we have a spatially uniform ground state with fully-polarized
spins even though the system is modulated. The energy spectrum
keeps this structure until $V_0$ reaches the critical value, where
the two Landau bands begin to touch each other at the energy of
$\mu = -\epsilon_c/2$. When $V_0$ is larger than the critical
value so that the bands overlap, the population of the spin-down
band becomes nonzero, and the spin splitting is no longer a
constant for different $X$'s [Fig.~\ref{fig2}(b)]. Thus a
partially polarized CDW state is formed, where both the charge and
the spin densities are periodically distributed. If $V_0$ is now
decreased, the energy spectrum will keep similar structure as that
in Fig.~\ref{fig2}(b). Only when $V_0$ is sufficiently small, such
that the difference of the populations between the spin-up and
spin-down states exceeds a critical amount, will the spin gap be
abruptly amplified. Hence the energy spectrum again becomes
similar to Fig.~\ref{fig2}(a) and the fully polarized state is
recovered. This history-dependent evolution of energy spectra is
in fact implied in Fig.~\ref{fig1}.

\begin{figure}
\centerline{\epsfxsize=7cm \epsfbox{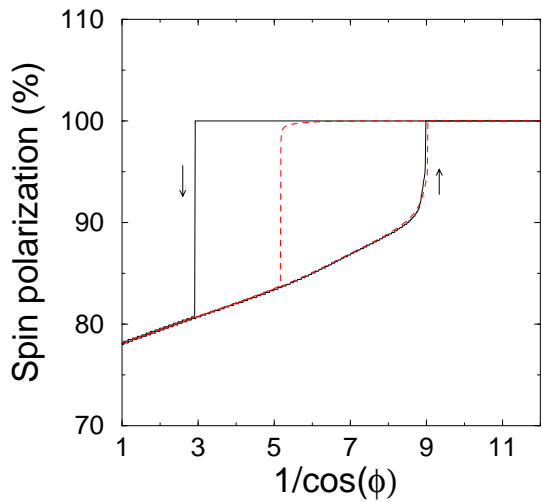} }
\caption{Hysteresis loops for the spin polarization for $k_B T=0$
(solid line) and $k_B T=0.01$ (dashed line). }\label{fig3}
\end{figure}

Instead of changing the modulation amplitude, the hysteresis loop
can also appear by tilting the magnetic field. For a given large
modulation amplitude (say, $V_0 = 0.6$), when the external
magnetic field is tilted by an angle $\phi$ but with a fixed
normal component $B_0$,
the total field $B=B_0/\cos\phi$ and therefore the Zeeman splitting
$\Delta_z=\Delta_z^{(0)}/\cos\phi$ are varied. By using the same
numerical iterative procedure, we show in Fig.~\ref{fig3} the spin
polarization for two temperatures ($k_B T=0$ and $k_B T=0.01$) with
$a=10$ and $V_0 = 0.6$ to illustrate the effect of thermal
fluctuation.
We find that, just as conventional ferromagnets, the hysteresis loop
shrinks as the temperature rises.

In conclusion, within the HFA and the LLL approximation, we
investigate hysteresis properties for the $\nu=1$ QH systems by
tuning the periodic modulation amplitude and the tilted magnetic
field. These results indicate that the intra-Landau-level exchange
interaction plays a crucial role in forming the hysteresis. This
hysteresis can be tested experimentally for modulated QH systems.
For typical GaAs-based samples with a carrier density about $3
\times 10^{11}$ cm$^{-2}$, the period and the amplitude of the
modulation required to exhibit the proposed hysteresis are about
0.1 $\mu$m and 10 meV, which can be reached by using current
technology. Although we focus our attention on the $\nu=1$ case
only, it is expected that similar hysteresis phenomena may occur
in some other fully spin-polarized incompressible QH states under
a periodic modulation, as long as the intra-Landau-level exchange
interaction plays an important role in forming the spin splitting.
For example, according to the composite-fermion theory,\cite{Jain}
the physics of the $\nu=1/3$ fractional QH state should be similar
to that of the composite fermion filling factor $\nu_{\rm CF}=1$
QH state. Therefore, by translating the present discussion into
the language of composite fermions, our results indicate that the
$\nu=1/3$ fractional QH state can be another candidate for this
modulation-induced hysteresis.

\acknowledgments
M.C.C. was supported by the National Science Council
of Taiwan under Contract No. NSC 89-2112-M-003-028. M.F.Y.
acknowledges financial support by the National Science Council of
Taiwan under Contract No. NSC 89-2112-M-029-006.

\widetext


\begin{thebibliography}{}


\bibitem[*]{} email address: changmc@phy03.phy.ntnu.edu.tw

\bibitem[\dagger]{} email address: mfyang@mail.thu.edu.tw

\bibitem{reviews}
{\em Perspectives in Quantum Hall Effect}, edited by S. Das Sarma and
A. Pinczuk (Wiley, New York, 1997).

\bibitem{Cho_et_al}
H. Cho {\it et al.}, Phys. Rev. Lett. {\bf 81}, 2522 (1998); H.~Cho
{\it et al.}, Physica {\bf E 6}, 18 (2000); J.~Eom {\it et al.},
Science, {\bf 289}, 2320 (2000).

\bibitem{Piazza_et_al}
V. Piazza {\it et al.}, Nature {\bf 402}, 638 (1999); V.~Piazza {\it
et al.}, Physica {\bf E 6}, 108 (2000). A similar hystersis in the
longitudinal and Hall resistances is also reported in a double-layer
2D {\it hole} gas at {\it even-integer} filling factors. See J. G. S.
Lok {\it et al.}, cond-mat/9907002.


\bibitem{Jungwirth_et_al}
T. Jungwirth {\it et al.}, Phys. Rev. Lett. {\bf 81}, 2328 (1998);
T.~Jungwirth and A.~H.~MacDonald, cond-mat/0003430.

\bibitem{modulation1}
R. R. Gerhardts, D. Weiss, and K. v. Klitzing, Phys. Rev. Lett. {\bf
62}, 1173 (1989); P. Vasilopoulos and F. M. Peeters, {\it ibid.} {\bf
63}, 2120 (1989).

\bibitem{modulation2}
D.~R.~Hofstadter, Phys.~Rev. {\bf 14}, 2239 (1976); M.~C.~Chang
and Q.~Niu, Phys. Rev. Lett. {\bf 75}, 1348 (1995).

\bibitem{Manolescu99:5426}
A. Manolescu and V. Gudmundsson, Phys. Rev. B {\bf 59}, 5426 (1999).

\bibitem{Manolescu00:7858}
A. Manolescu and V. Gudmundsson, Phys. Rev. B {\bf 61}, R7858 (2000).

\bibitem{Ando}
T. Ando and Y. Uemura, J. Phys. Soc. Jpn. {\bf 37}, 1044 (1974).


\bibitem{Bychkov}
Yu. A. Bychkov {\it et al.}, Phys. Rev. Lett. {\bf 73}, 2911 (1994);
Europhys. Lett. {\bf 40}, 557 (1997). T. Maniv {\it et al.}, Physica
{\bf B 204}, 134 (1995).

\bibitem{Cote}
R. C\^{o}t\'{e} and A. H. MacDonald, Phys. Rev. Lett. {\bf 65}, 2662
(1990); Phys. Rev. B {\bf 44}, 8759 (1991). R.~C\^{o}t\'{e} and H. A.
Fertig, {\it ibid.} {\bf 62}, 1993 (2000).

\bibitem{note2}
However, the estimate based on the HFA will be somewhat larger
than that obtained in Ref.~\protect\onlinecite{Bychkov}, since the
vertex correction of the Coulomb interaction, which accounts for
the attraction between the electron and the hole of a spin
exciton, is neglected in the HFA. For example, for the system
parameters used in Ref.~\protect\onlinecite{Bychkov} ($G_0 =0.4$
and $\Delta_z^{(0)} =0.03$), our critical value of $V_0$ is 0.538,
while the value obtained in Ref.~\protect\onlinecite{Bychkov} is
0.465.

\bibitem{Jain}
For example, see J. K. Jain in Ref.~\protect\onlinecite{reviews}.

\end{thebibliography}
\end{document}